\documentclass[journal,draftclsnofoot,onecolumn,12pt,twoside]{IEEEtranTCOM}
\normalsize

\usepackage{pgfplots}

\ifCLASSINFOpdf
\else
\fi

\usepackage{amssymb}
\usepackage{amsthm}
\usepackage[cmex10]{amsmath}
\usepackage{algorithm}
\usepackage{algorithmic}
\usepackage{array}
\usepackage{mdwmath}
\usepackage{mdwtab}
\usepackage[tight,footnotesize]{subfigure}
\usepackage{url}
\usepackage[center]{caption}
\usepackage{float}
\usepackage{subfigure}
\usepackage{epsfig,mathptm, verbatim}
\usepackage{graphicx}
\usepackage{amssymb}
\usepackage{tikz}
\usepackage{hyperref}
\usepackage{breakurl}
\usepackage{amsbsy}
\renewcommand{\vec}[1]{\mathbf{#1}}
\newcommand{\sfrac}[2]{\raise0.1ex\hbox{$#1$} \! \! \left/ \! \lower0.6ex\hbox{$#2$}\right.}
\newcommand{\sfracsmall}[2]{\raise0.1ex\hbox{$#1$} \! \! \left/ \! \lower0.6ex\hbox{$#2$}\right.}

\newenvironment{definition}[1][Definition]{\begin{trivlist}
\item[\hskip \labelsep {\underline{#1}:}]}{\end{trivlist}}

\theoremstyle{plain}
    \newtheorem{theorem}{Theorem}
    \newtheorem{corollary}{Corollary}[theorem]
    
	\newtheorem{remark}{Remark}

    \newtheoremstyle{TheoremNum}
        {\topsep}{\topsep}              
        {\itshape}                      
        {}                              
        {\bfseries}                     
        {.}                             
        { }                             
        {\thmname{#1}\thmnote{ \bfseries #3}}
    \theoremstyle{TheoremNum}

\usepackage{cite}
\usepackage[square, comma, sort&compress,numbers]{natbib}
\newcolumntype{L}[1]{>{\raggedright\let\newline\\\arraybackslash\hspace{0pt}}m{#1}}
\newcolumntype{C}[1]{>{\centering\let\newline\\\arraybackslash\hspace{0pt}}m{#1}}
\newcolumntype{R}[1]{>{\raggedleft\let\newline\\\arraybackslash\hspace{0pt}}m{#1}}
\usepackage{multirow}

\date{}

\usepackage{lipsum}

  \hypersetup{linktocpage} 
  \hypersetup{
    colorlinks,
    citecolor=blue,
    filecolor=black,
    linkcolor=red,
    urlcolor=black
}

\floatname{algorithm}{Procedure}

\begin{document}
\title{Blind Fractional Interference Alignment}
\author{B Hari Ram \hspace{4mm} G Kanchana Vaishnavi \hspace{4mm} K Giridhar \\Department of Electrical Engineering \\ Indian Institute of Technology Madras \\ Chennai-600036, India\\Email:[hariram, giri]@tenet.res.in, vaishnavigandikota@gmail.com}
\maketitle

\begin{abstract}
Fractional Interference Alignment (FIA) is a transmission scheme which achieves any value between $[0,1]$ for the Symbols transmitted per Antenna per Channel use (SpAC). 
FIA was designed in \cite{HariRam2013b} specifically for Finite Alphabet (FA) signals, under the constraint that the Minimum Distance (MD) detector is used at all the receivers. Similar to classical interference alignment, the FIA precoder also needs perfect channel state information at all the transmitters (CSIT). 
In this work, a novel Blind Fractional Interference Alignment (B-FIA) scheme is introduced, where the basic assumption is that CSIT is not available. We consider two popular channel models, namely: Broadcast channel, and Interference channel. For these two channel models, the maximum achievable value of SpAC satisfying the constraints of the MD detector is obtained, but with no CSIT, and also a precoder design is provided to obtain any value of SpAC in the achievable range. 

Further, the precoder structure provided has one distinct advantage: interference channel state information at the receiver (I-CSIR) is not needed, when all the transmitters and receivers are equipped with one antenna each. When two or  more antennas are used at both ends, I-CSIR must be available to obtain the maximum achievable value of SpAC. The receiver designs for both the Minimum Distance and the Maximum Likelihood (ML) decoders are discussed, where the interference statistics is estimated from the received signal samples. 
Simulation results of the B-FIA show that the ML decoder with estimated statistics achieves a significantly better error rate performance when compared to the MD decoder with known statistics,
since the MD decoder assumes the interference plus noise term as colored Gaussian noise. 
\end{abstract}

\begin{IEEEkeywords}
Interference Alignment, Fractional Interference Alignment, Blind Fractional Interference Alignment, Blind Interference Alignment, Finite Alphabet Signals, Non-Linear Receiver, $K-$user Interference Channel, Symbols transmitted per transmit-Antenna per Channel use (SpAC).
\end{IEEEkeywords}

\section{Introduction}\label{sec:Intro}
Co-channel interference is generally considered as a hindrance while improving the throughput of a $K-$user wireless network.
In 2007, Jafar and Shamai introduced the concept of Interference Alignment (IA) in a $2-$user X-channel \cite{Jafar2007a,Jafar2008} and later, Cadambe and Jafar \cite{Cadambe2008} applied the IA concept in a $K-$user Interference Channel (IC) and showed that each user gets $\sfrac{M}{2}$ Degrees-of-Freedom (DoF) irrespective of the number of the interfering signals seen by the receiver, where $M$ represents the number of dimensions available for transmission. In \cite{Nazer2009, Nazer2012} ergodic IA was used to achieve $\sfrac{M}{2}$ DoF. However, the IA schemes in \cite{Cadambe2008,Nazer2009, Nazer2012} require perfect channel knowledge of all the desired and interfering channels at each transmitter (or equivalently, requires a centralized controller which designs the precoders, and sends them respectively to each transmitter).

In practice, it is very difficult to implement the conventional IA schemes \cite{Jafar2007a,Jafar2008,Cadambe2008,Nazer2009, Nazer2012} because of the requirement of perfect channel knowledge. Hence in \cite{Jafar2010} and \cite{Jafar2012}, Jafar introduced a Blind IA (B-IA) scheme which exploits the correlation present in the channel, i.e., channel should have specific states to achieve the alignment of interference since Channel State Information at the Transmitter~(CSIT) is not available. In \cite{Gou2011}, the channel states required for B-IA to work were achieved through a staggered antenna switching scheme at the receiver. In \cite{Maleki2012}, a retrospective IA scheme was introduced which uses delayed CSIT to achieve the alignment of interfering signals.

Furthermore, practical communication links use Finite Alphabet (FA) constellations such as $M-$QAM signals, whereas the IA schemes \cite{Jafar2007a,Jafar2008,Cadambe2008,Nazer2009, Nazer2012,Jafar2010,Jafar2012,Gou2011,Maleki2012} assume the usage of Gaussian signals at each transmitter. Hence, we proposed in \cite{HariRam2013b} the Fractional Interference Alignment (FIA) scheme where the alignment of interference is based on new set of constraints (as briefly discussed in section \ref{sec:FIAOverview}), which explicitly considers the usage of FA signals at all the transmitters. A new metric named SpAC\footnote{SpAC represents the number of message streams sent per transmit-Antenna per Channel use. Unlike DoF, the SpAC metric is a more general expression, since the optimum value (for achieving better bit error rate, or mutual information, or both) of SpAC will be a function of both signal-to-noise ratio and signal-to-interference ratio. } was introduced \cite{HariRam2013b}, and it was shown that using the 
FIA scheme any value of SpAC in the full range $[0,1]$ can be achieved. Numerical results were provided in \cite{HariRam2013c} where the FIA precoder design was compared with the existing IA precoder design, and was shown to achieve a better BER performance. Once again, like the IA scheme, the FIA scheme also requires accurate CSIT to attain the maximum SpAC value of $1$ per transmitter, which is practically not feasible.

Hence, in our current work, a novel Blind Fractional Interference Alignment (B-FIA)\footnote{B-FIA is a `blind transmission' scheme like the B-IA, where the transmitters does not have CSIT.} scheme is introduced, based on the following two practical considerations or constraints: (i)~FA signals are used at all the transmitters, and (ii)~CSIT is not available. Like the FIA scheme, the B-FIA scheme is a collection of all precoder designs which satisfy the constraints, $\mathcal{S}^{Rx}\nsubseteq\mathcal{I}^{Rx}$, but without  CSIT, where $\mathcal{S}^{Rx}$ and $\mathcal{I}^{Rx}$ represent the desired signal subspace and interfering signal subspace at the receiver, respectively. Also, B-FIA can be viewed as an extension of the B-IA scheme in \cite{Jafar2010} and \cite{Jafar2012}, with the additional practical choice of FA transmit signals. However, there is another key difference between the B-IA and B-FIA schemes: B-IA requires specific channel states to achieve the DoF \cite{Jafar2012}, while the proposed B-
FIA does not impose any such unrealistic condition on the channel states. In this work, it will be analytically shown that the maximum achievable value of SpAC, for a Multi Input Multi Output (MIMO) Broadcast and Interfering channels, is $\sfrac{(L-K\frac{M}{N}+1)}{L}$, when $M\leq N$. 
It must be mentioned that the maximum achievable value of SpAC for the case when $M > N$ is still an open problem. Here, $M$ represents the number of transmitter antennas, $N$ represents the number of receiver antennas, and $L$ represents the symbol extension factor\footnote{The number of resource elements used for precoding is known as the symbol extension factor, since the effect of one symbol is present across many symbol durations.}. We also propose a simple precoder design which achieves this maximum value of SpAC, and satisfies the required constraint,~$\mathcal{S}^{Rx}\nsubseteq\mathcal{I}^{Rx}$.

The proposed precoder structure has an advantage for Single Input Single Output (SISO) channels. This advantage is that the statistics\footnote{Statistics can be either: probability density function, or the covariance matrix, required for the ML detector, or the MD detector, respectively.} of the interfering signal can be estimated from the received data signals, i.e., no external controls or pilot signals are required to estimate the statistics of the interfering signal. Hence, we term this precoder design as Complete B-FIA (CB-FIA)\footnote{CB-FIA is a `blind detection' scheme for the proposed precoder design. Unlike classical unsupervised learning or adaptive estimation literature where `blind' signifies the absence of training data or pilots, here `blind' is used to indicate non-availability of CSIT. Hence, we have used `complete blind' to indicate that I-CSIR is also not required.} design where Interference Channel State Information at Receiver (I-CSIR) \footnote{I-CSIR implies either interference 
covariance or the joint probability distribution of the interfering signals is known at the receiver. In either case the desired channel condition is known at the receivers. Throughout this work this definition of I-CSIR is followed.} is not required. However, the CB-FIA scheme can achieve the maximum value of SpAC only for SISO links \footnote{
When the CB-FIA is extended to MIMO and SIMO links, we speculate that the statistics of the interfering signals can be estimated only when MIMO or SIMO are employed with a SpAC of $\sfrac{(L-K+1)}{L}$.}. 
A general channel model is considered while obtaining the maximum achievable SpAC in the B-FIA scheme, and it cannot be compared with B-IA scheme (which requires specific channel states \cite{Jafar2010}). Hence, in this work, we do not compare the proposed B-FIA scheme with the B-IA scheme.

We also discuss two detectors for the CB-FIA scheme, namely: (i)~Maximum Likelihood (ML) detector, and (ii)~Minimum Distance (MD) detector. The estimation procedure required for obtaining the statistics of the interfering signals is also discussed.

In summary, we introduce the B-FIA scheme which is the collection of precoder designs that satisfies the constraint  $\mathcal{S}^{Rx} \nsubseteq \mathcal{I}^{Rx}$. Analytically, we show that the maximum achievable value of SpAC is $\sfrac{(L-K\frac{M}{N}+1)}{L}$, and introduce a simple precoder design to achieve the maximum value of SpAC. For SISO, the proposed precoder design has an inherent advantage of not requiring the I-CSIR, and hence we term it as the \text{CB-FIA} scheme. Finally, we compare the bit error rate (BER) performance of the ML and MD detectors considered for this B-FIA scheme.



\subsection*{Notation:} 
Span($\vec{A}$) is the column space of the matrix $\vec{A}$, $|\text{span}(\vec{A})|$ is the dimension of the column space of $\vec{A}$ or the rank of the matrix $\vec{A}$, with some abuse of notation we represent the $|\text{span}(\vec{A})|$ by $|\vec{A}|$, and $\text{I}_M$ represents the $M\times M$ identity matrix. Further, $\mathcal{X}_i$ represents the set containing all possible values of the transmitted symbol vector $\vec{x}_i$ and the elements are assumed to be ordered, $\vec{x}_{i,j}$ represents the $j^{th}$ vector element of the set $\mathcal{X}_i$, {$d_i$} is the collection of all distance metric at $i^{th}$ receiver, and $d_i^{[jk]}$ represents the distance between $\vec{x}_{i,j}$ and $\vec{x}_{i,k}$ ($\vec{x}_{i,j}, \vec{x}_{i,k} \in \mathcal{X}_i$), at the $i^{th}$ receiver. Finally, $\vec{e}_i$ represents the axis vector whose ith element is unity and all other elements are zero. 

\section{Channel Model and Overview of B-IA} \label{sec:ChanMod}
\subsection{Broadcast Channel} \label{sec:BC}
A $K-$user BC model consists of one transmitter with $M$ antennas and $K-$receivers with $N$ antennas each. The considered scenario is BC with multiple multicast messages, i.e., the single transmitter transmits independent data signals to each receiver. The received signal at the $i^{th}$ receiver is given~by,
\begin{equation}
\begin{array}{lll}
\vec{y}_i = \vec{H}_{i,i} \sum_{j=1}^K \vec{Q}_j \vec{x}_j + \vec{n}_i
\end{array},
\end{equation}
where
\begin{itemize}
\item $\vec{y}_i$ \--- received signal at $i^{th}$ receiver of dimension $NL\times 1$,
\item $\vec{x}_j$ \--- transmitted signal intended for $j^{th}$ receiver of dimension $d\times 1$,
\item $\vec{n}_i$ \--- Complex Gaussian noise, $\vec{n}_i\in \mathcal{CN}(\vec{0},\sigma^2 \vec{I}_{NL})$,
\item $\vec{H}_{i,i}$ \--- channel between the transmitter and $i^{th}$ receiver of dimension $NL\times ML$,
\item $\vec{Q}_j$ \--- precoder matrix for $\vec{x}_j$ of dimension $ML\times d$,
\item $L$ \--- symbol extension factor.
\end{itemize}
In \cite{Jafar2010}, it was shown that in the $2-$user BC with no CSIT, the achievable DoF is $\sfrac{4}{3}$ when the transmitter is equipped with $2$ antennas, and each receiver is equipped with $1$ antenna. This DoF was achieved for a specific channel state obtained by exploiting the correlation present in the channel response across time and frequency. The channel structure seen at the $2$ receivers, and the corresponding precoder matrices are given by, \cite{Jafar2010},
\begin{gather}\label{eqn:BIAchannel}
\begin{array}{llll}
\vec{H}_{1,1}=\begin{bmatrix}
\vec{h}_{1,1}(1)^T & \vec{0}_{1\times 2} & \vec{0}_{1\times 2}\\
\vec{0}_{1\times 2} & \vec{h}_{1,1}(2)^T & \vec{0}_{1\times 2}\\
\vec{0}_{1\times 2} & \vec{0}_{1\times 2} & \vec{h}_{1,1}(1)^T 
\end{bmatrix}, \;\; \vec{Q}_1 = \begin{bmatrix}
\vec{I}_2\\ \vec{I}_2 \\ \vec{0}_{2\times 2}
\end{bmatrix}\\
\text{and,}\\
\vec{H}_{2,2}=\begin{bmatrix}
\vec{h}_{2,2}(1)^T & \vec{0}_{1\times 2} & \vec{0}_{1\times 2}\\
\vec{0}_{1\times 2} & \vec{h}_{2,2}(1)^T & \vec{0}_{1\times 2}\\
\vec{0}_{1\times 2} & \vec{0}_{1\times 2} & \vec{h}_{2,2}(2)^T 
\end{bmatrix}, \;\; \vec{Q}_2 = \begin{bmatrix}
\vec{I}_2 \\ \vec{0}_{2\times 2} \\ \vec{I}_2
\end{bmatrix}
\end{array}\raisetag{3.7\baselineskip}
\end{gather}
The result in \cite{Gou2011} is an extension of the B-IA in \cite{Jafar2010} for a $K-$user BC, where the required channel state is achieved by assuming that each receiver is equipped with $M$ antennas. Of the $M$ receive antennas, only one antenna is used at each time slot, where the method of antenna selection \cite{Gou2011} guarantees B-IA. And, B-IA can achieve $\sfrac{MK}{(M+K-1)}$ DoF. 
We first describe the B-IA scheme in the interference channel before discussing the proposed B-FIA scheme.

\subsection{Interference Channel} \label{sec:IC}
A $K-$user IC model consists of $K$ transmitters with $M$ antennas each and $K-$receivers with $N$ antennas each. Each transmitter has a useful message to only one receiver, and each receiver receives useful message from only one transmitter. The presence of other transmitters results in $K-1$ interfering terms getting added to the desired signal at each of the $K$ receivers. The received signal at the $i^{th}$ receiver is given by,
\begin{equation}\label{eqn:IC_chanMod}
\begin{array}{lll}
\vec{y}_i = \sum_{j=1}^K \vec{H}_{i,j} \vec{Q}_j \vec{x}_j + \vec{n}_i
\end{array},
\end{equation}
where
$\vec{H}_{i,j}$ represents the channel between the $j^{th}$ transmitter and $i^{th}$ receiver of dimension $NL\times ML$,
$\vec{Q}_j$ is the precoder matrix for $\vec{x}_j$ of dimension $ML\times d$,
and $L$ is the symbol extension factor.

In \cite{Jafar2010}, it was shown that with no CSIT, the achievable DoF is $\sfrac{K}{2}$, provided that all the desired channel gains $\vec{H}_{i,i}$ are time varying within one block (of $L$ symbols), while all the interfering channels $\vec{H}_{i,j,i \neq j}$ are constant within that block. 

The B-IA scheme for both BC and IC scenarios assumes the use of a linear receiver; i.e., the precoder design constraints are based on linear detection. However, when FA signals are used, the optimum receiver is based on either the MAP or ML (Maximum A-Posteriori or Maximum Likelihood) principle, and in practical scenarios, BER will be the performance measure. Hence, SpAC was introduced in \cite{HariRam2013b} to show the advantage of the FIA scheme when FA signals are used. 
The value of SpAC for BC and IC is $\sfrac{MK}{(M+K-1)}$ and $\sfrac{K}{2}$ respectively, using the B-EIA\footnote{We call B-IA scheme with FA signals as Blind Extended IA (B-EIA) scheme, since we are interested only when FA signals are used.} scheme. However, the \text{B-EIA} scheme requires the special channel states \eqref{eqn:BIAchannel} which might not occur in a practical scenario. Also, as mentioned earlier there is no constraint on the channel matrices for B-FIA scheme. Hence, we do not compare the SpAC of the B-FIA scheme with the B-EIA scheme.
In this work, we will obtain the maximum achievable value of SpAC for the B-FIA scheme for both BC and IC where the precoder design constraints are based on the non-linear Minimum Distance (MD) detector \cite{Kuchi2011}. We also provide a simple precoder design to achieve that value of SpAC. 

\section{Overview of Fractional Interference Alignment} \label{sec:FIAOverview}
The FIA scheme was introduced for the scenario where all the transmitters uses FA signals, and in such a scenario, the optimal receiver is a non-linear receiver. The precoder design constraint for the FIA scheme is given by \cite{Kuchi2011},
\begin{equation}\label{eqn:FIAConst}
\begin{array}{lll}
\mathcal{S}_i \nsubseteq \mathcal{I}_i
\end{array},
\end{equation}
where $\mathcal{S}_i$, and $\mathcal{I}_i$ represent the subspace occupied by the desired signal and the interfering signal, respectively, at the $i^{th}$ receiver. For a $K-$user IC, the constraint (\ref{eqn:FIAConst}) can be rewritten as,
\begin{subequations}\label{eqn:FIAConst_KIC}
\begin{align}
&|\bigcup\limits_{j=1,j\neq i}^K \text{span}(\vec{H}_{i,j} \vec{Q}_j)| < NL\\
\text{and, }& |\bigcup\limits_{j=1}^K \text{span}(\vec{H}_{i,j} \vec{Q}_j)| = NL
\end{align}
\end{subequations}
Using the constraints in (\ref{eqn:FIAConst_KIC}), it was shown in \cite{HariRam2013b} that any value of SpAC in the full range $[0,1]$ is asymptotically achievable, for $K-$user IC when CSIT is available. In a practical scenario, it is very difficult to make available accurate CSIT available. Each receiver can usually estimate the desired channel $\vec{H}_{i,i}$, but estimating (and feeding back) the interfering channels $\vec{H}_{i,j}$ might not be feasible unless special pilot structures are available for the same. 
Hence, the question to be answered is,
\textit{`With no CSIT, what is the maximum achievable value of SpAC?'}

It will be shown in section \ref{sec:BFIAPrecoders} that the SpAC given in Table~\ref{tab:Different_SpACBCIC} is the maximum value achievable when MD detector is used.

\begin{table}[h]
\centering
\ifCLASSOPTIONtwocolumn
\scalebox{1.0}{
\fi
\begin{tabular}{|c|c|c|}
\hline 
& BC & IC \\
\hline
SISO&$\sfrac{(L-K+1)}{L}$&$\sfrac{(L-K+1)}{L}$\\[0.1 cm]
\hline
MIMO&$\sfrac{(ML-K+1)}{ML}$&$\sfrac{(ML-K\frac{M}{N}+1)}{ML}$\\
\hline
\end{tabular}
\ifCLASSOPTIONtwocolumn
}
\fi
\caption{Maximum achievable SpAC and hence the achievable Range of SpAC} \label{tab:Different_SpACBCIC}
\end{table}


\section{Blind Fractional Interference Alignment} \label{sec:BFIAPrecoders}
The design criteria for the Blind Fractional Interference Alignment scheme is influenced by: (i)~the usage of FA signal set at each transmitter, (ii)~the availability of an appropriate non-linear receiver, and (iii)~the non-availability of CSIT at each transmitter (or equivalently, the absence of a centralized controller). In this section, we will provide the maximum achievable value of SpAC in SISO, SIMO and MIMO (with $M\leq N$) scenarios for both BC and IC.

It must be mentioned here that all the proofs provided in this section assume the number of transmit antenna is less than the number of receive antenna. For MIMO with $M>N$, the number of dimensions each sub-space occupies is still limited by the number of receiver antennas, $N$, i.e., $|\vec{H}_{i,i}\vec{Q}_i| = d\leq NL$, which is equivalent to having $N$ antennas at the transmitter. Hence, the results obtained will hold only for the case when the transmitters are equipped with $N$ antennas. As mentioned earlier, for $M>N$ (MISO is a special case with $N=1$ and $M>1$), the maximum achievable value of SpAC is still an open problem.


In the next two sub-sections (section \ref{subsec:BC_B-FIA} and \ref{subsec:IC_B-FIA}), we will show that the SpAC values given in Table \ref{tab:Different_SpACBCIC} are the maximum values for the respective scenarios. In section~\ref{subsec:PrecStr}, we will discuss about one possible precoder structure to achieve these maximum value of SpAC, and this design can be extended to attain any achievable value of SpAC.

\subsection{Broadcast Channel}\label{subsec:BC_B-FIA}
\begin{theorem}\label{Thm:SISOBC}
The maximum achievable SpAC in a Broadcast Channel with no CSIT for SISO case is $\sfrac{(L-K+1)}{L}$.
\end{theorem}

\textit{Proof:} Consider a $K-$user BC with single antenna at the transmitter and at each receiver, with symbol extension factor of $L$. Since it is a BC, and the channel gains of the desired signal are known at all the receivers, the signal at $l^{th}$ receiver is given by,
\begin{equation}\label{app:eqn:BCChanMod}
\begin{array}{lll}
\vec{y}_l = \sum\limits_{i=1}^K \vec{Q}_i \vec{x}_i + \vec{n}_l
\end{array}.
\end{equation}

From (\ref{eqn:FIAConst}), the constraints are given by,
\begin{subequations}\label{app:eqn:consSISOBC}
\begin{align}
&|\bigcup\limits_{i=1}^K \vec{Q}_i| = L \label{app:eqn:consSISOBC_a}\\
\text{and, }& |\bigcap\limits_{\substack{j\subset \{1,\cdots,K\},\\|j|>1}} \vec{Q_j}| < d \label{app:eqn:consSISOBC_b}
\end{align}
\end{subequations}
where $|\vec{Q}_i|=d$. To prove Theorem \ref{Thm:SISOBC}, we need to show that the maximum value of $d$ satisfying (\ref{app:eqn:consSISOBC}) is $L-K+1$. We know that,
\ifCLASSOPTIONtwocolumn
\begin{equation}\label{app:eqn:unions}
\begin{array}{llll}
|\bigcup\limits_{i=1}^K \vec{Q}_i| = \sum\limits_{i_1=1}^K \underbrace{|\vec{Q}_{i_1}|}_{d} - \sum\limits_{i_1=1}^K\sum\limits_{i_2=i_1+1}^K \underbrace{|\vec{Q}_{i_1} \cap \vec{Q}_{i_2}|}_{d-\epsilon_{i_1 i_2}} \\
\hspace{15mm}+ \sum\limits_{i_1=1}^K\sum\limits_{i_2=i_1+1}^K \sum\limits_{i_3=i_2+1}^K \underbrace{|\vec{Q}_{i_1} \cap \vec{Q}_{i_2} \cap \vec{Q}_{i_3}|}_{d-\epsilon_{i_1 i_2 i_3}} - \cdots \\
\hspace{15mm}+ (-1)^{K+1} \sum\limits_{i_1=1}^K \sum\limits_{i_2=i_1+1}^K \cdots \sum\limits_{i_K=i_{K-1}+1}^K \underbrace{|\bigcap\limits_{j=i_1}^{i_K} \vec{Q_j}|}_{d-\epsilon_{i_1\cdot i_K}}
\end{array}
\end{equation}
\begin{equation}
\begin{array}{l}
L = K d - {K\choose 2}d + {K\choose 3} d -\cdots + (-1)^K {K\choose K} d +  \epsilon'
\end{array}.
\end{equation}
\else
\begin{equation}\label{app:eqn:unions}
\begin{array}{clll}
|\bigcup\limits_{i=1}^K \vec{Q}_i| = \sum\limits_{i_1=1}^K \underbrace{|\vec{Q}_{i_1}|}_{d} - \sum\limits_{i_1=1}^K\sum\limits_{i_2=i_1+1}^K \underbrace{|\vec{Q}_{i_1} \cap \vec{Q}_{i_2}|}_{d-\epsilon_{i_1 i_2}} + \sum\limits_{i_1=1}^K\sum\limits_{i_2=i_1+1}^K \sum\limits_{i_3=i_2+1}^K \underbrace{|\vec{Q}_{i_1} \cap \vec{Q}_{i_2} \cap \vec{Q}_{i_3}|}_{d-\epsilon_{i_1 i_2 i_3}} - \cdots\\
\hspace{70mm} + (-1)^{K+1} \sum\limits_{i_1=1}^K \sum\limits_{i_2=i_1+1}^K \cdots \sum\limits_{i_K=i_{K-1}+1}^K \underbrace{|\bigcap\limits_{j=i_1}^{i_K} \vec{Q_j}|}_{d-\epsilon_{i_1\cdot i_K}}
\end{array}
\end{equation}
\begin{equation}
\begin{array}{l}
L = K d - {K\choose 2}d + {K\choose 3} d -\cdots + (-1)^K {K\choose K} d +  \epsilon'
\end{array},
\end{equation}
\fi
where $\epsilon_{i\cdot l}$ denotes the number of column spaces in $\vec{Q}_j, j=1\cdot l$ which does not overlap with the column spaces of the other precoder matrices. 
Hence,
\begin{equation}
\begin{array}{lll}
d = L - \epsilon'
\end{array},
\end{equation}
where $\epsilon'$ is given by,
\ifCLASSOPTIONtwocolumn
\begin{equation}
\begin{array}{lll}
\hspace{-1mm}\epsilon' = -\sum\limits_{i_1=1}^K\sum\limits_{i_2=i_1+1}^K \epsilon_{i_1 i_2} + \sum\limits_{i_1=1}^K\sum\limits_{i_2=i_1+1}^K \sum\limits_{i_3=i_2+1}^K \epsilon_{i_1 i_2 i_3} - \cdots \\
\hspace{17mm}+ (-1)^{K+1}\sum\limits_{i_1=1}^K \sum\limits_{i_2=i_1+1}^K \cdots \sum\limits_{i_K=i_{K-1}+1}^K \epsilon_{i_1\cdot i_K}
\end{array}
\end{equation}
\else
\begin{equation}
\begin{array}{lll}
\hspace{-1mm}\epsilon' = -\sum\limits_{i_1=1}^K\sum\limits_{i_2=i_1+1}^K \epsilon_{i_1 i_2} + \sum\limits_{i_1=1}^K\sum\limits_{i_2=i_1+1}^K \sum\limits_{i_3=i_2+1}^K \epsilon_{i_1 i_2 i_3} - \cdots 
+ (-1)^{K+1}\sum\limits_{i_1=1}^K \sum\limits_{i_2=i_1+1}^K \cdots \sum\limits_{i_K=i_{K-1}+1}^K \epsilon_{i_1\cdot i_K}
\end{array}
\end{equation}
\fi
For maximizing the SpAC the intersection of sub-spaces 
$d-\epsilon_{i\cdot l}$
should be maximized. Hence, all the $\epsilon$'s should be made as minimum as possible. Thus all $\epsilon_{i\cdot l}$'s will be set equal to $1$ ($\because$~$\epsilon_{i\cdot l} > 0$, and $\epsilon_{i\cdot l} \in \mathbb{Z}$), and the value of $\epsilon'$ will be $K-1$. Hence, the maximum value of $d$ is $(L-K+1)$, which completes the proof for Theorem~\ref{Thm:SISOBC}.

\begin{corollary}\label{lem:MIMOBC}
The maximum achievable SpAC in a Broadcast Channel with no CSIT for MIMO ($M\leq N$) link is $\sfrac{(ML-K+1)}{ML}$.
\end{corollary}


\textit{Proof:} Premultiplying the received signal vector by a full rank matrix results in modifying the sub-space occupied by the desired and interfering signal at each receiver. However, for the $M=N$ case, the dimension of the sub-space formed by the union and intersection of desired and interfering signal will not change. Hence, when $M=N$, pre-multiplying the $i^{th}$ received signal by $\vec{H}_{i,i}^{-1}$ will result in SISO IC with the symbol extension factor increased from $L$ to $ML$. 
Using Theorem~\ref{Thm:SISOBC}, the maximum achievable value of SpAC is $\sfrac{(ML-K+1)}{ML}$. Since the considered scenario is a BC, increasing the number of antennas at receiver will not provide any advantage in spatial dimensions for the desired signal over the interfering signal. Hence, the maximum achievable value of SpAC for $N>M$ is also $\sfrac{(ML-K+1)}{ML}$, which completes the proof.

\begin{corollary}\label{lem:SIMOBC}
The maximum achievable SpAC in a Broadcast Channel with no CSIT for SIMO link is $\sfrac{(L-K+1)}{L}$.
\end{corollary}

\textit{Proof:} This is a special case of MIMO BC with $M=1$.

\begin{remark}
{\em
If the channel states have some special structure (as in B-IA, \cite{Jafar2010} or (\ref{eqn:BIAchannel})) it might be possible to achieve higher value of SpAC, but those channel states are highly unlikely to occur in a practical scenario. Hence, such special channel states as in (\ref{eqn:BIAchannel}), are not considered in our work.
}
\end{remark}


\subsection{Interference Channel}\label{subsec:IC_B-FIA}
\begin{theorem}\label{Thm:SIMOIC}
The maximum achievable SpAC in a Interference Channel with no CSIT for SIMO link is $\sfrac{(L-\frac{K}{N}+1)}{L}$.
\end{theorem}

\textit{Proof:} Consider a $K-$user IC with single antenna at each transmitter and $N$ antennas at each receiver, with symbol extension factor of $L$. The received signal at $l^{th}$ receiver is given by,
\begin{equation}\label{app:eqn:ICChanMod}
\begin{array}{lll}
\vec{y}_l = \sum\limits_{i=1}^K \vec{H}_{l,i} \vec{Q}_i \vec{x}_i + \vec{n}_l
\end{array},
\end{equation}
where $\vec{y}_l$ is a $NL\times 1$ vector and $\vec{H}_{l,i}$ is a $NL \times L$ block diagonal matrix, $\vec{x}_i$ is a $d\times 1$ message signal vector, and $\vec{Q}_i$ is the $L\times d$ precoder matrix at $i^{th}$ transmitter. The proof for Theorem \ref{Thm:SIMOIC} is obtained by finding the maximum value of $d$ which satisfies~(\ref{eqn:FIAConst}). From (\ref{eqn:FIAConst}), the constraint equations can also be given by,
\begin{subequations}\label{app:eqn:consSIMOIC}
\begin{align}
&|\bigcup\limits_{i=1}^K \vec{H}_{l,i} \vec{Q}_i| = NL \label{app:eqn:consSIMOIC_a}\\
\text{and, }& |(\bigcup\limits_{i=1,i\neq l}^K \vec{H}_{l,i} \vec{Q}_i)\; \cap\; \vec{H}_{l,l}\vec{Q}_l| < d \label{app:eqn:consSIMOIC_b}
\end{align}
\end{subequations}
Now,
\ifCLASSOPTIONonecolumn
\begin{equation}\label{app:eqn:unionICexp}
\begin{array}{llll}
|\bigcup\limits_{i=1}^K \vec{H}_{l,i} \vec{Q}_i| &=& |\bigcup\limits_{i=1}^{K-1} \vec{H}_{l,i} \vec{Q}_i| + |\vec{H}_{l,K} \vec{Q}_K| - |(\bigcup\limits_{i=1}^{K-1} \vec{H}_{l,i} \vec{Q}_i) \;\cap \; \vec{H}_{l,K} \vec{Q}_K|\\
|\bigcup\limits_{i=1}^K \vec{H}_{l,i} \vec{Q}_i|&\operatorname*{=}\limits^{(a)}& \sum\limits_{i=1}^K |\vec{H}_{l,i} \vec{Q}_i| - \sum\limits_{i=2}^K |(\bigcup\limits_{j=1}^{j<i} \vec{H}_{l,j} \vec{Q}_j) \;\cap \; \vec{H}_{l,i} \vec{Q}_i|
\end{array}
\end{equation}
\else
\begin{equation}\label{app:eqn:unionICexp}
\begin{array}{llll}
|\bigcup\limits_{i=1}^K \vec{H}_{l,i} \vec{Q}_i| &=& |\bigcup\limits_{i=1}^{K-1} \vec{H}_{l,i} \vec{Q}_i| + |\vec{H}_{l,K} \vec{Q}_K|\\
&&\hspace{1cm} - |(\bigcup\limits_{i=1}^{K-1} \vec{H}_{l,i} \vec{Q}_i) \;\cap \; \vec{H}_{l,K} \vec{Q}_K|\\
|\bigcup\limits_{i=1}^K \vec{H}_{l,i} \vec{Q}_i|&\operatorname*{=}\limits^{(a)}& \sum\limits_{i=1}^K |\vec{H}_{l,i} \vec{Q}_i| - \sum\limits_{i=2}^K |(\bigcup\limits_{j=1}^{j<i} \vec{H}_{l,j} \vec{Q}_j) \;\cap \; \vec{H}_{l,i} \vec{Q}_i|
\end{array}
\end{equation}
\fi
where $(a)$ is obtained by expanding the $|\bigcup\; (\cdot)|$, recursively.
The constraints in (\ref{app:eqn:consSIMOIC}) are true for all $l={1,\cdots, K}$. Since CSIT is not available, the constraint (\ref{app:eqn:consSIMOIC_b}) can also be rewritten as,
\begin{equation}\label{app:eqn:consSIMOIC_b_mod}
\begin{array}{lll}
|(\bigcup\limits_{i=1,i\neq j}^K \vec{H}_{l,i} \vec{Q}_i)\; \cap\; \vec{H}_{l,j}\vec{Q}_j| = d - \epsilon_{lj}
\end{array}.
\end{equation}
Since the channels are independent, and the number of transmit antennas is $1$, we get,
\begin{equation}\label{app:eqn:consSIMOIC_b_mod_0}
\begin{array}{lll}
|(\bigcup\limits_{\substack{i\subset {1,\cdots,K},\\i\neq j,|i|<N}} \vec{H}_{l,i} \vec{Q}_i)\; \cap\; \vec{H}_{l,j}\vec{Q}_j| = 0
\end{array}.
\end{equation}
Using (\ref{app:eqn:consSIMOIC_b_mod}) and (\ref{app:eqn:consSIMOIC_b_mod_0}) in (\ref{app:eqn:unionICexp}), we get,
\begin{equation}
\begin{array}{lll}
|\bigcup\limits_{i=1}^K \vec{H}_{l,i} \vec{Q}_i|&= \sum\limits_{i=1}^K |\vec{H}_{l,i} \vec{Q}_i| - \sum\limits_{i=N+1}^K |(\bigcup\limits_{j=1}^{j<i} \vec{H}_{l,j} \vec{Q}_j) \;\cap \; \vec{H}_{l,i} \vec{Q}_i|\\
&= Kd - \sum\limits_{i=N+1}^K (d-\epsilon_{l,i})\\
&= Nd + \sum\limits_{i=N+1}^K (\epsilon_{l,i})
\end{array}.
\end{equation}
Once again, for maximizing the SpAC, the intersection of the sub-spaces in (\ref{app:eqn:consSIMOIC_b_mod}) should be maximized. Thus all the $\epsilon$'s should be made as minimum as possible. As in the proof for Theorem \ref{Thm:SISOBC}, all the values of $\epsilon$'s will be set equal to~$1$. Hence,
\begin{equation}
\begin{array}{lll}
NL = Nd + (K-N)\\
d = L - \sfrac{K}{N} +1
\end{array},
\end{equation}
which completes the proof for Theorem \ref{Thm:SIMOIC}.

\begin{corollary}\label{lem:MIMOIC}
The maximum achievable SpAC in an Interference Channel with no CSIT for MIMO link is $\sfrac{(ML-\frac{MK}{N}+1)}{ML}$.
\end{corollary}

\textit{Proof:} The proof for corollary \ref{lem:MIMOIC} follows the proof for Theorem \ref{Thm:SIMOIC} with the only difference being that (\ref{app:eqn:consSIMOIC_b_mod_0}) is slightly modified when there are two or more antennas at each transmitter, and is given by,
\begin{equation}\label{app:eqn:consMIMOIC_b_mod_0}
\begin{array}{lll}
|(\bigcup\limits_{\substack{i\subset {1,\cdots,K},\\i\neq j,|i|<\frac{N}{M}}} \vec{H}_{l,i} \vec{Q}_i)\; \cap\; \vec{H}_{l,j}\vec{Q}_j| = 0
\end{array}.
\end{equation}
Using (\ref{app:eqn:consMIMOIC_b_mod_0}) in place of (\ref{app:eqn:consSIMOIC_b_mod_0}) in (\ref{app:eqn:unionICexp}) will give the maximum value of $d$ as $ML - \sfrac{KM}{N} + 1$, which completes the proof for corollary \ref{lem:MIMOIC}.

\begin{corollary}\label{lem:SISOIC}
The maximum achievable SpAC in an Interference Channel with no CSIT for SISO link is $\sfrac{(L-K+1)}{L}$.
\end{corollary}

\textit{Proof:} This is a special case of SIMO IC with $N=1$.

\subsection{Precoder Structure to achieve Maximum SpAC}\label{subsec:PrecStr}
In the previous two sub-sections (section \ref{subsec:BC_B-FIA} and \ref{subsec:IC_B-FIA}), the maximum achievable value of SpAC was obtained for both BC and IC models. In this sub-section, we will provide the structure of precoder matrices which achieve the maximum possible value of SpAC.

Even though the above proofs do not give an explicit form for the precoder, the maximum achievable value of SpAC provides the intuition on the structure of the precoder. For example, the SpAC value of $\sfrac{(L-K+1)}{L}$ for SISO implies: (i)~the symbols are precoded over $L$ symbols, and (ii)~number of symbols transmitted is $L-K+1$. In other words, each transmitter sends ($K-1$) symbols less than the maximum value of $L$. Each receiver receives ($K-1$) interfering signals, and along with the constraint that there should exist a sub-space in $\mathcal{S}$ where no interfering signal is present (i.e., $\mathcal{S} \nsubseteq \mathcal{I}$). This leads to the simple design rule given in Procedure \ref{Proc:PrecDesign} to obtain the maximum achievable SpAC.
\begin{algorithm}[h]
\flushleft
\ifCLASSOPTIONtwocolumn
\begin{minipage}{0.5\textwidth}
\else
\begin{minipage}{1\textwidth}
\fi
\caption{Design Rule for B-FIA to achieve $\sfrac{d}{L}$~SpAC}
\label{Proc:PrecDesign}
\begin{enumerate}
\item Separate the available resources ($L$) into two groups
\begin{enumerate}
\item $K$ resources for the $1^{st}$ group.
\item $L-K$ resources for the $2^{nd}$ group.
\end{enumerate}
\item From the $1^{st}$ group, assign one resource to each transmitter.
\item From the $2^{nd}$ group, assign some $d-1$ resources to each transmitter. The presence of $1^{st}$ group will ensure the existence of a subspace in $\mathcal{S}$ where no interfering is signal present.
\item Finally, use any arbitrary (or for a better performance, optimized) set of $d\times d$ precoder matrices, $\vec{U}$, across those $d$ resources at each transmitter.
\end{enumerate}
\ifCLASSOPTIONtwocolumn
\end{minipage}
\else
\end{minipage}
\fi
\end{algorithm}

For a $3-$user SISO channel, the precoder structure that will achieve the maximum value of SpAC with $L=4$ is given~by,
\ifCLASSOPTIONtwocolumn
\begin{gather} \label{eqn:ICBM_Matrices}
\begin{array}{ccc}
\vec{Q}_1 = \begin{bmatrix}
1 & 0 \\
0 & 0 \\
0 & 0 \\
0 & 1 \\
\end{bmatrix} \vec{U}_1,\;\;
\vec{Q}_2 = \begin{bmatrix}
0 & 0 \\
1 & 0 \\
0 & 0 \\
0 & 1 \\
\end{bmatrix} \vec{U}_2,
\vec{Q}_3 = \begin{bmatrix}
0 & 0 \\
0 & 0 \\
1 & 0 \\
0 & 1 \\
\end{bmatrix} \vec{U}_3
\end{array}.\raisetag{1\baselineskip}
\end{gather}
\else
\begin{gather} \label{eqn:ICBM_Matrices}
\begin{array}{ccc}
\vec{Q}_1 = \begin{bmatrix}
1 & 0 \\[-0.3cm]
0 & 0 \\[-0.3cm]
0 & 0 \\[-0.3cm]
0 & 1 \\[-0.3cm]
\end{bmatrix} \vec{U}_1,\;\;
\quad \vec{Q}_2 = \begin{bmatrix}
0 & 0 \\[-0.3cm]
1 & 0 \\[-0.3cm]
0 & 0 \\[-0.3cm]
0 & 1 \\[-0.3cm]
\end{bmatrix} \vec{U}_2,
\quad \vec{Q}_3 = \begin{bmatrix}
0 & 0 \\[-0.3cm]
0 & 0 \\[-0.3cm]
1 & 0 \\[-0.3cm]
0 & 1 \\[-0.3cm]
\end{bmatrix} \vec{U}_3
\end{array}.\raisetag{1\baselineskip}
\end{gather}
\fi
For any unitary matrix $\vec{U}_i$\footnote{The unitary matrices are introduced to satisfy the necessary condition, $\vec{H_{i,i}} \vec{q}_i^{[j]} \nsubseteq \vec{I}$, where $\vec{q}_i^{[j]}$ represent the $j^{th}$ column vector of the matrix $\vec{Q}_i$.}$,\;i=1,2,3$, this structure will satisfy constraint~(\ref{eqn:FIAConst}), and CSIT is not required. Hence, this design is termed as Blind Fractional Interference Alignment scheme, where the term `blind' indicates CSIT is not available, and the term `fractional' is used because any fractional value for SpAC in the range $[0,\sfrac{(L-K+1)}{L}]$ can be achieved with this precoder structure\footnote{Throughout this work when (\ref{eqn:ICBM_Matrices}) is referred, it implies the extended version of this structure for the appropriate values of $L$ and $K$.}. A similar procedure can be followed for SIMO and MIMO scenarios to obtain the precoder matrices.

\section{Decoder for B-FIA}\label{sec:Dec_BFIA}
As mentioned earlier, the precoder structure (\ref{eqn:ICBM_Matrices}) has an advantage for SISO IC, even while achieving the maximum possible SpAC. This advantage is that the corresponding decoder does not require the parameters of the interfering signals, and in this work we will term it as no interfering signal's CSIR (I-CSIR) required.
Of course, the receivers require accurate estimate of the desired channel gains, and it is assumed that pilot signals are available to estimate the same. 
In this section, we describe about two types of detector for the B-FIA scheme: Minimum Distance (MD) and Maximum Likelihood (ML) detectors. The procedure for estimating the parameters of the interfering signals, for the respective detectors is also briefly discussed.
\subsection{Completely Blind FIA (No I-CSIR)}
\subsubsection{ML Detector}
For finite alphabet transmissions, we use the Gaussian Mixture Model (GMM) to model the probability density function (pdf) of the interference plus noise, as given in \cite{Yang2013}. 
The desired signal can be decoded from the received signal vector (\ref{app:eqn:ICChanMod}), using Maximum likelihood (ML) detection as follows,
\begin{equation}\label{eqn:GMM_detect}
\begin{array}{llll}
\hat{\vec{x}}_i = \underset{\vec{x}_i\in \mathcal{X}_i}{\operatorname{argmax}}\;\; p_{\vec{y}_i|\vec{g}_i}(\vec{y}_i|\vec{g}_i,\vec{x}_i)
\end{array},
\end{equation}
where $p_{\vec{y}_i|\vec{g}_i}(\vec{y}_i|\vec{g}_i,\vec{x}_i)$ is the conditional joint pdf of the received signal, and $\vec{g}_i$ is the parameter set that characterizes the received signal from transmitter $i$, with $\vec{g}_i=[\mathcal{X}_{i},\vec{H}_{i,i},\vec{Q}_i]$. 
The structure provided in (\ref{eqn:ICBM_Matrices})  can be useful in estimating the interference plus noise pdf from the received signal at each receiver.
Consider the signal model, $\vec{z} \in \mathbf{C}^{m\times 1}$, given by
\begin{equation} \label{eqn:InterMod}
\begin{array}{lll}
\vec{z} = \vec{U} \vec{x} + \vec{n}
\end{array},
\end{equation}
where
$\vec{U}$ is a Unitary matrix of dimension $m\times m$, and $\vec{n}$ is a complex Gaussian vector, $\vec{n}\in \mathcal{CN}(\vec{0},\sigma^2 \vec{I}_{m})$.


\begin{theorem} \label{Thm:samePDF}
For any \underline{useful} real unitary matrix with even number of columns (i.e., $m=2l,\;\forall l\in \mathbb{N}$), the elements of
 $\vec{z}$ will be identically distributed, if~$\vec{x}~\in~\mathcal{X}^m$ and $\mathcal{X} = \{\pm a_i,\; i=1,\cdots, \sfrac{|\mathcal{X}|}{2}, \;a_i\in\mathbb{C} \}$.
\end{theorem}
\textit{Proof:} Please refer Appendix \ref{App:ThmsamePDF_Pf}.

The term \textit{useful} real unitary matrix has been defined
in Appendix \ref{App:ThmsamePDF_Pf}. When the real unitary matrix is obtained from an optimization problem in which all the symbols transmitted have equal priors, then the real unitary matrix can be written as a function of $(m-1)$ variables, and we refer to this as an \textit{useful} real unitary matrix.

FA signal sets like $P-$PSK (for even $P$), $P^2-$QAM, symmetric lattice etc., has the form of the set $\mathcal{X}$. 
\textit{Thus, estimating the marginal pdf of one of the element of $\vec{z}$, results in the estimation of marginal pdf of all the elements of $\vec{z}$.}

Even though the joint distribution of $\vec{z}$ cannot be formed without the knowledge of $\vec{U}$ and $\mathcal{X}$, having only the marginal pdf alone will be useful in certain scenarios (including the design of the B-FIA, as will be explained later in this section).\\


\noindent \underline{\textbf{Estimation:}}
For a block fading channel, the received signal at the $i^{th}$ receiver for a SISO channel with symbol extension factor of $L$ is given as,
\begin{equation}\label{eqn:IC_chan_blocSISO}
\begin{array}{lll}
\vec{y}_i &=& \sum_{j=1}^K h_{i,j} \vec{Q}_j \vec{x}_j + \vec{n}_i\\
&=& h_{i,i} \vec{Q}_i \vec{x}_i + \vec{z}_i
\end{array},
\end{equation}
where $\vec{z}_i$ represent the interference plus noise at the $i^{th}$ receiver, and $\vec{y}_i\in \mathbf{C}^{L\times 1}$.
For the above mentioned model, the procedure to obtain the joint pdf  $p_{\vec{y}_i|\vec{g}_i}(\vec{y}_i|\vec{g}_i,\vec{x}_i)$ as in (\ref{eqn:GMM_detect}) will be explained for the $2-$user, and the $3-$user SISO IC, and the extension to $K-$user SISO IC then becomes straightforward.\\

\noindent \textbf{$2-$user SISO IC with $L=3$:}
With precoder as in (\ref{eqn:ICBM_Matrices}), the interference plus noise at the first receiver is given by,
\begin{equation}\label{eqn:2userSISO_int}
\begin{array}{llll}
\vec{z}_1 = \begin{bmatrix}
0 & 0\\
1 & 0\\
0 & 1
\end{bmatrix} h_{1,2} \vec{U}_2 \vec{x}_2 + \vec{n}_1
\end{array},
\end{equation}
and the received signal is given by,
\begin{equation}\label{eqn:2userSISO_Rx1}
\begin{array}{lll}
\begin{bmatrix}
y_1^{[1]} \\
y_1^{[2]} \\
y_1^{[3]}
\end{bmatrix} = \begin{bmatrix}
h_{1,1}\vec{u}_1^{[1]T}\vec{x}_1 + n_1^{[1]}\\
0 + z_1^{[2]}\\
h_{1,1}\vec{u}_1^{[2]T}\vec{x}_1 + z_1^{[3]}
\end{bmatrix}
\end{array},
\end{equation}
where $a^{[j]}$ (or $\vec{a}^{[j]T}$) represent the $j^{th}$ row of the vector (or matrix) $\vec{A}$. From (\ref{eqn:2userSISO_Rx1}), the marginal pdf of $z_1^{[2]}$ can be estimated using the Expectation-Maximization (EM) algorithm (as given in \cite{Redner1984}) from the received signal samples, $y_1^{[2]}$. Hence, using Theorem \ref{Thm:samePDF}, the marginal pdf of $z_1^{[3]}$ is also known, and is given by,
\begin{equation}\label{eqn:2userSISO_margpdf}
\begin{array}{llll}
p_{z_1}(z_1) &=& \frac{1}{|\mathcal{X}_2|} \sum\limits_{\vec{x}_2 \in \mathcal{X}_2} \frac{1}{\pi \sigma^2} exp(-\frac{\lVert z_1 - h_{1,2}\vec{u}_2^{[1]T} \vec{x}_2  \rVert^2}{\sigma^2})
\end{array}.
\end{equation}
In (\ref{eqn:2userSISO_margpdf}), we used $z_1$ instead of $z_1^{[j]}$ ($\forall j \neq 1$) since the marginal pdfs are the same from Theorem~\ref{Thm:samePDF}. 
Finding $p(\vec{y}_1|\vec{g}_1,\vec{x}_1)$ requires the knowledge of $\vec{Q}_2$, and because of $\vec{Q}_2$, $p(y_1^{[2]}|\vec{g}_1,\vec{x}_1)$ and $p(y_1^{[3]}|\vec{g}_1,\vec{x}_1)$ are dependent. However in CB-FIA, as $\vec{Q}_2$ is assumed to be unknown at the first receiver, we have considered only the elements of $\vec{y}_1$ in which the desired signal is present.
Now, the (sub-optimal) ML detector (\ref{eqn:GMM_detect}) becomes,
\begin{equation}\label{eqn:2userSISO_GMM_detect}
\begin{array}{llll}
\hspace{-3mm} \hat{\vec{x}}_1 &=& \underset{\vec{x}_1\in \mathcal{X}_1}{\operatorname{argmax}}\;\; p_{\vec{y}_1|\vec{g}_1}(y_1^{[1]}|\vec{g}_1,\vec{x}_1) \;\; p_{\vec{y}_1|\vec{g}_1}(y_1^{[3]}|\vec{g}_1,\vec{x}_1) \\
&=& \underset{\vec{x}_1\in \mathcal{X}_1}{\operatorname{argmax}}\;\; p_{n_1}(y_1^{[1]}-h_{1,1}\vec{u}_1^{[1]T}\vec{x}_1) \;\; p_{z_1}(y_1^{[3]}-h_{1,1}\vec{u}_1^{[2]T}\vec{x}_1)
\end{array},
\end{equation}
where $p_{n_1}(\cdot)$ represents the marginal pdf of the additive white Gaussian noise. In (\ref{eqn:2userSISO_GMM_detect}), the joint pdf is written as the product of two marginal pdfs, because $z_1$ and $n_1$ are independent.
A similar procedure will be followed for the second user.\\

\noindent \textbf{$3-$user SISO IC with $L=4$:}
As given for the $2-$user SISO IC, the interference plus noise signal for the $3-$user case, at the first receiver is given by,
\ifCLASSOPTIONtwocolumn
\begin{equation}\label{eqn:3userSISO_int}
\begin{array}{llll}
\hspace{-1mm}\vec{z}_{1,2} = \begin{bmatrix}
0 & 0\\
1 & 0\\
0 & 0\\
0 & 1
\end{bmatrix} h_{1,2} \vec{U}_2 \vec{x}_2 + \vec{n}_1, \;\;
\vec{z}_{1,3} = \begin{bmatrix}
0 & 0\\
0 & 0\\
1 & 0\\
0 & 1
\end{bmatrix} h_{1,3} \vec{U}_3 \vec{x}_3 + \vec{n}_1,\\
\hspace{-1mm}\text{and, }\hspace{15mm}
\vec{z}_{1} = \vec{z}_{1,2} + \vec{z}_{1,3} - \vec{n}_1,
\end{array}
\end{equation}
\else
\begin{equation}\label{eqn:3userSISO_int}
\begin{array}{llll}
\hspace{-1mm}\vec{z}_{1,2} = \begin{bmatrix}
0 & 0\\[-0.3cm]
1 & 0\\[-0.3cm]
0 & 0\\[-0.3cm]
0 & 1\\[-0.3cm]
\end{bmatrix} h_{1,2} \vec{U}_2 \vec{x}_2 + \vec{n}_1, \;\;
\qquad \vec{z}_{1,3} = \begin{bmatrix}
0 & 0\\[-0.3cm]
0 & 0\\[-0.3cm]
1 & 0\\[-0.3cm]
0 & 1\\[-0.3cm]
\end{bmatrix} h_{1,3} \vec{U}_3 \vec{x}_3 + \vec{n}_1,\\
\hspace{-10mm}\text{and, }\hspace{30mm}
\vec{z}_{1} = \vec{z}_{1,2} + \vec{z}_{1,3} - \vec{n}_1,
\end{array}
\end{equation}
\fi
and the received signal is given by,
\begin{equation}\label{eqn:3userSISO_Rx1}
\begin{array}{lll}
\begin{bmatrix}
y_1^{[1]} \\
y_1^{[2]} \\
y_1^{[3]} \\
y_1^{[4]} 
\end{bmatrix} = \begin{bmatrix}
h_{1,1}\vec{u}_1^{[1]T}\vec{x}_1 + n_1^{[1]}\\
0 + z_{1,2}^{[2]}\\
0 + z_{1,3}^{[3]}\\
h_{1,1}\vec{u}_1^{[2]T}\vec{x}_1 + z_1^{[4]}
\end{bmatrix}
\end{array},
\end{equation}
where $z_1^{[4]}$ represent the sum of all the interfering signals along with the noise. Similar to the $2-$user SISO IC, the pdf of $z_{1,2}^{[2]}$, and $z_{1,3}^{[3]}$ can be estimated from the received signal samples $y_{1}^{[2]}$, and $y_{1}^{[3]}$, respectively, using the EM algorithm. However, the marginal pdf of the individual interference plus noise term cannot be directly used as in $2-$user SISO IC. The marginal pdf of $z_1^{[4]}$ has to be constructed from the marginal pdf of $z_{1,2}^{[2]}$, and $z_{1,3}^{[3]}$. 
Both $z_{1,2}^{[2]}$, and $z_{1,3}^{[3]}$ have a Gaussian mixture pdf, with variance of all components being equal to $\sigma^2$, and means given by $\{h_{1,2} \vec{u}_2^{[1]T} \vec{x}_2\}$, and $\{h_{1,3} \vec{u}_3^{[1]T} \vec{x}_3\}$, respectively. Now, $z_1^{[4]}$ also has a Gaussian mixture pdf whose means can be obtained using $*$\footnote{Here, $\{a\}*\{b\}$ denotes the set $\{a_i + b_j,\;\; \forall\; i,j \}$, and $a_i$ and $b_j$ represents the element of the sets $\{a\}$ and $\{b\}$.} as $\{\{h_{1,2}\vec{u}_2^{[1]T} \vec{x}_2\}~*~\{h_{1,3}\vec{u}_3^{[1]T} \vec{x}_3\}\}$, and variance of all the components equal $\sigma^2$. 
Hence, the pdf of the interference plus noise term is given by,
\begin{equation}\label{eqn:3userSISO_margpdf}
\begin{array}{llll}
\hspace{-1mm} p_{z_1}(z_1) = c \sum\limits_{\vec{x}_2 \in \mathcal{X}_2} \sum\limits_{\vec{x}_3 \in \mathcal{X}_3}  exp(-\frac{\lVert z_1 - h_{1,2}\vec{u}_2^{[1]T} \vec{x}_2 - h_{1,3}\vec{u}_3^{[1]T} \vec{x}_3 \rVert^2}{\sigma^2})
\end{array},
\end{equation}
where $c=\frac{1}{|\mathcal{X}_2||\mathcal{X}_3|} \frac{1}{(\pi \sigma^2)^2}$. Thus, considering only the elements of $\vec{y}_1$ in which desired signal is present, the ML detector (\ref{eqn:GMM_detect}) becomes, 
\begin{equation}\label{eqn:3userSISO_GMM_detect}
\begin{array}{llll}
\hspace{-3mm} \hat{\vec{x}}_1 
= \underset{\vec{x}_1\in \mathcal{X}_1}{\operatorname{argmax}}\;\; p_{n_1}(y_1^{[1]}-h_{1,1}\vec{u}_1^{[1]T}\vec{x}_1) \;\; p_{z_1}(y_1^{[4]}-h_{1,1}\vec{u}_1^{[2]T}\vec{x}_1)
\end{array}.
\end{equation}

\noindent \textbf{$K-$user SISO IC with $L=K+1$:}
For $K-$user SISO IC with $L=K+1$, similar procedure can be followed, i.e., estimate the pdf of each interfering signal from the first group in the precoder design, and use it to construct the pdf of the total interference plus noise. Since the estimate of the interfering signal plus noise is known, the ML detector can be used to detect the desired signal.\\

\noindent \textbf{$K-$user SISO IC with $L>K+1$:}
The considered scenario so far was $d=2$. In the detectors formed using (\ref{eqn:2userSISO_GMM_detect}) and (\ref{eqn:3userSISO_GMM_detect}), the joint pdf was written as product of two marginal pdfs because: (i)~the noise is independent across each dimension, and (ii)~the number of dimension in which the interfering signal is present along with desired signal is only one. When $d>2$, the joint pdf cannot be obtained as the product of the marginal pdf, since the independence assumption is no longer valid. This is because the interfering signal is present along with the desired signal in more than one dimension (for maximum achievable SpAC, it is $d-1$ dimensions), and the interfering signal is not independent across each dimension\footnote{When a unitary matrix is used in the precoder, the interference is uncorrelated across each dimension, but not independent.}. However, extending the same procedure as in $d=2$, an approximate joint pdf can be obtained, even for $d>2$. \\

\subsubsection{MD detector}
If the interference plus noise is assumed to be coloured Gaussian, then the covariance of the interference plus noise can be 
estimated for the precoder structure in (\ref{eqn:ICBM_Matrices}) (please refer Appendix~\ref{App:covEst}). Using the covariance matrix, the desired signal can be decoded using the MD receiver \footnote{For estimating the covariance matrix, channel has to be block fading across estimation duration. Hence $\vec{H}_{i,i}$ will be replace by $h_{i,i}$.}\cite{Kuchi2011},
\begin{equation}\label{eqn:MD_detect}
\begin{array}{llll}
\hat{\vec{x}}_i = \underset{\vec{x}_i\in \mathcal{X}_i}{\operatorname{argmin}}\; \parallel\vec{y}_i-h_{i,i}\vec{Q}_i \vec{x}_i\parallel_{\vec{R}_i^{-1}}^2
\end{array},
\end{equation}
where $\vec{R}_i$ is the covariance of the interference plus noise term, and $\parallel \mathbf{a}\parallel_\mathbf{B}^2 = \mathbf{a}^\text{H} \mathbf{B} \mathbf{a}$.

 \begin{remark}
For MIMO IC, the pdf estimation for SISO channels can be extended, but with the SpAC value limited to that for the SISO model. Complete B-FIA for a MIMO IC with no I-CSIR is an open problem, to achieve the corresponding maximum value of SpAC.
 \end{remark}
 
 \begin{remark}
 For the existing single stream transmission in MISO IC like beamforming, maximal ratio transmission, etc, the CB-FIA can be employed with symbol extension, i.e., no external pilots are needed at the receiver while aligning the interfering signals. In such cases the multiple transmit antennas are treated as a single virtual transmit antenna, and the maximum achievable SpAC is $\sfrac{(L-K+1)}{L}$.
 \end{remark}

\subsection{CSIR available} \label{subsec:SpAC_CSIR_available}
When CSIR is available, either the ML decoder (\ref{eqn:GMM_detect}) or the MD decoder (\ref{eqn:MD_detect}) can be used to detect the desired signal. However, the constraint for the B-FIA is based on the MD detector~(\ref{eqn:MD_detect}). Hence, when the ML decoder is used, and with CSIR available, the maximum achievable SpAC will be more than the SpAC given in Table \ref{tab:Different_SpACBCIC}. In fact, with ML decoding and perfect CSIR available, the maximum achievable SpAC is unity.
This can be shown by, (i)~treating the ML decoder as the MD decoder, and (ii)~the constraint $\mathcal{S} \nsubseteq \mathcal{I}$. In the ML decoder, all the interfering signals can be decoded, so we treat all the signals as the desired signals, and the ML decoder as the MD decoder with the covariance matrix $\vec{R}_i$ contains only complex Gaussian noise. Thus, there is no interfering signal, or $\mathcal{I}=\emptyset$, and it will satisfy the constraint $\mathcal{S} \nsubseteq \mathcal{I}$. Hence, the ML decoder can achieve $1$ SpAC, and this was verified numerically in section~\ref{sec:NumRes}.

\section{Numerical Results}\label{sec:NumRes}
\begin{figure*}
\centering
\begin{minipage}{0.5\textwidth}
\centering
\includegraphics[scale = 0.5]{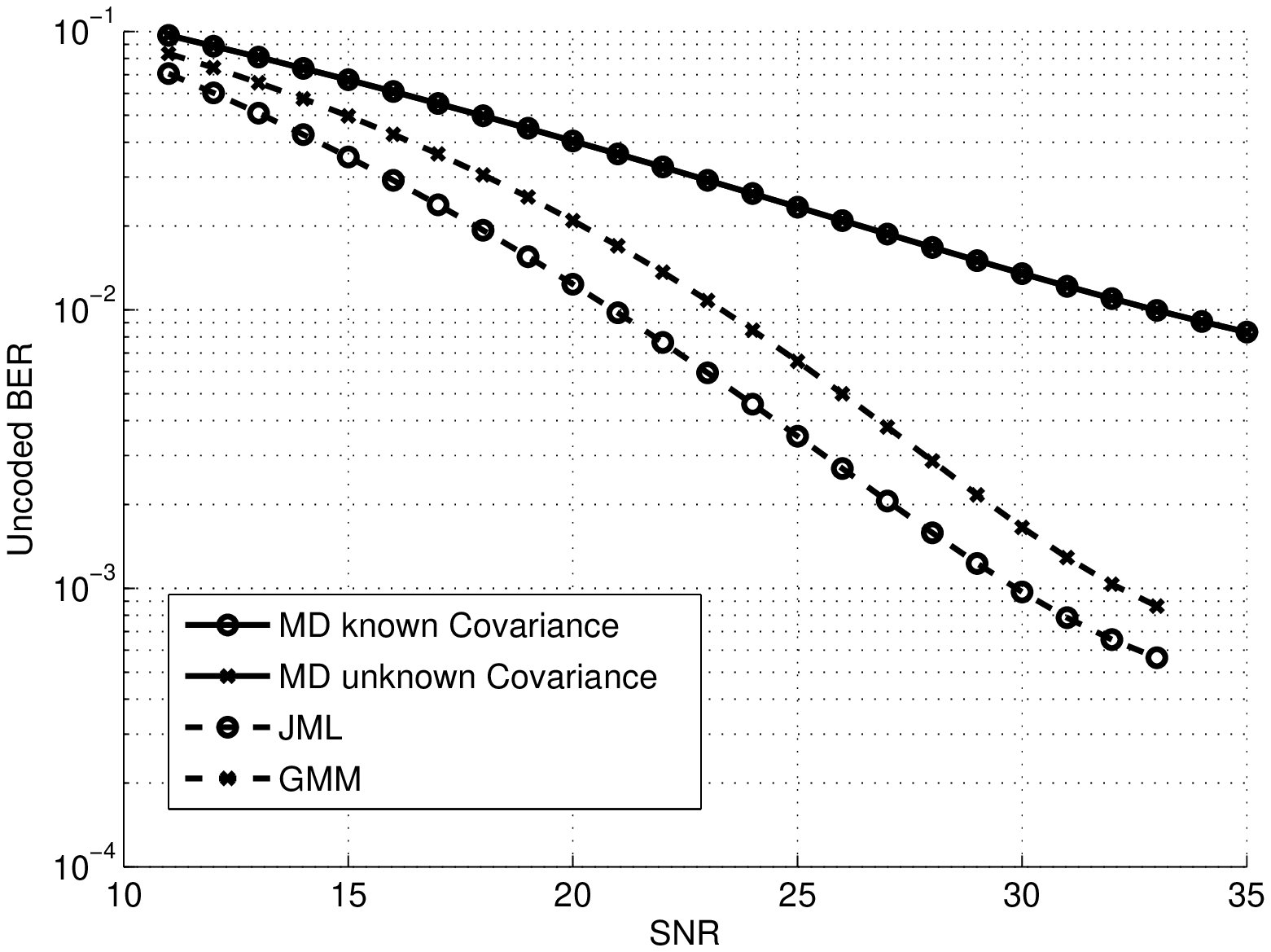}
\caption{Uncoded BER: $K=2,\; L=3,\; d=2$}
\label{fig:Un_K_2_M_3_d_2}
\end{minipage}%
\begin{minipage}{0.5\textwidth}
\centering
\includegraphics[scale = 0.5]{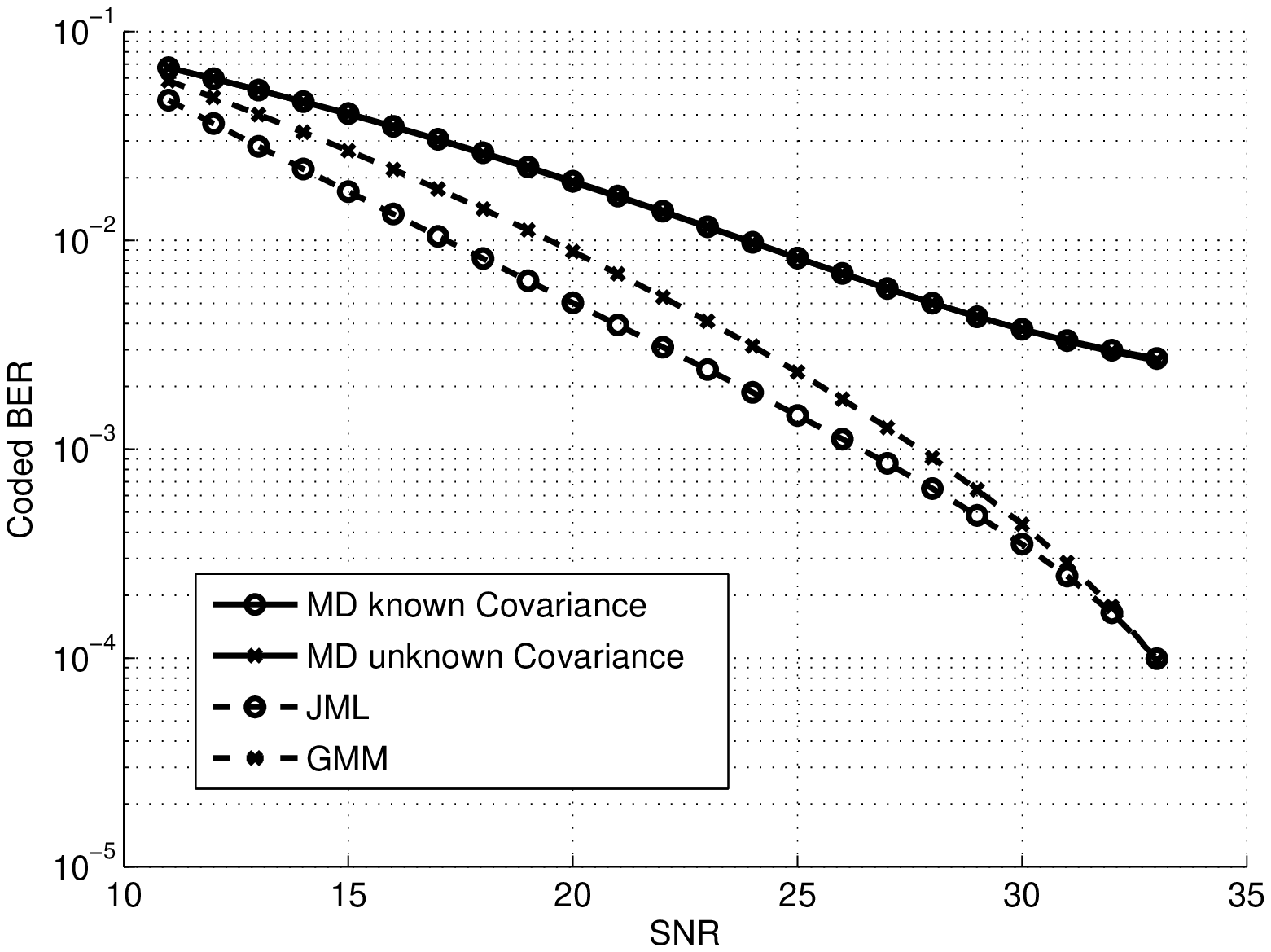}
\caption{Coded BER: $K=2,\;L=3,\;d=2$}
\label{fig:C_K_2_M_3_d_2}
\end{minipage}

\begin{minipage}{0.5\textwidth}
\centering
\includegraphics[scale = 0.5]{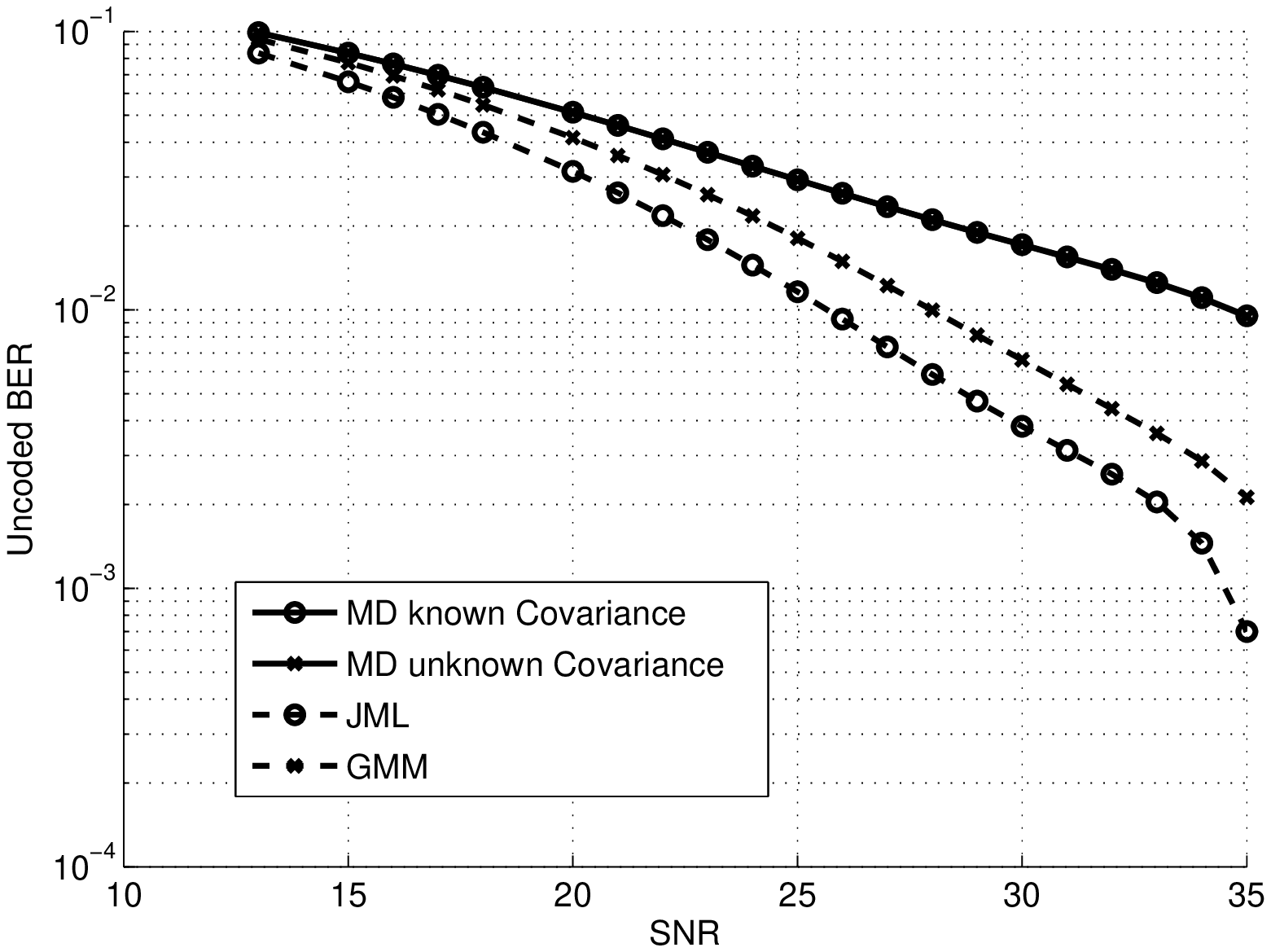}
\caption{Uncoded BER: $K=3,\;L=4,\;d=2$}
\label{fig:Un_K_3_M_4_d_2}
\end{minipage}%
\begin{minipage}{0.5\textwidth}
\centering
\includegraphics[scale = 0.5]{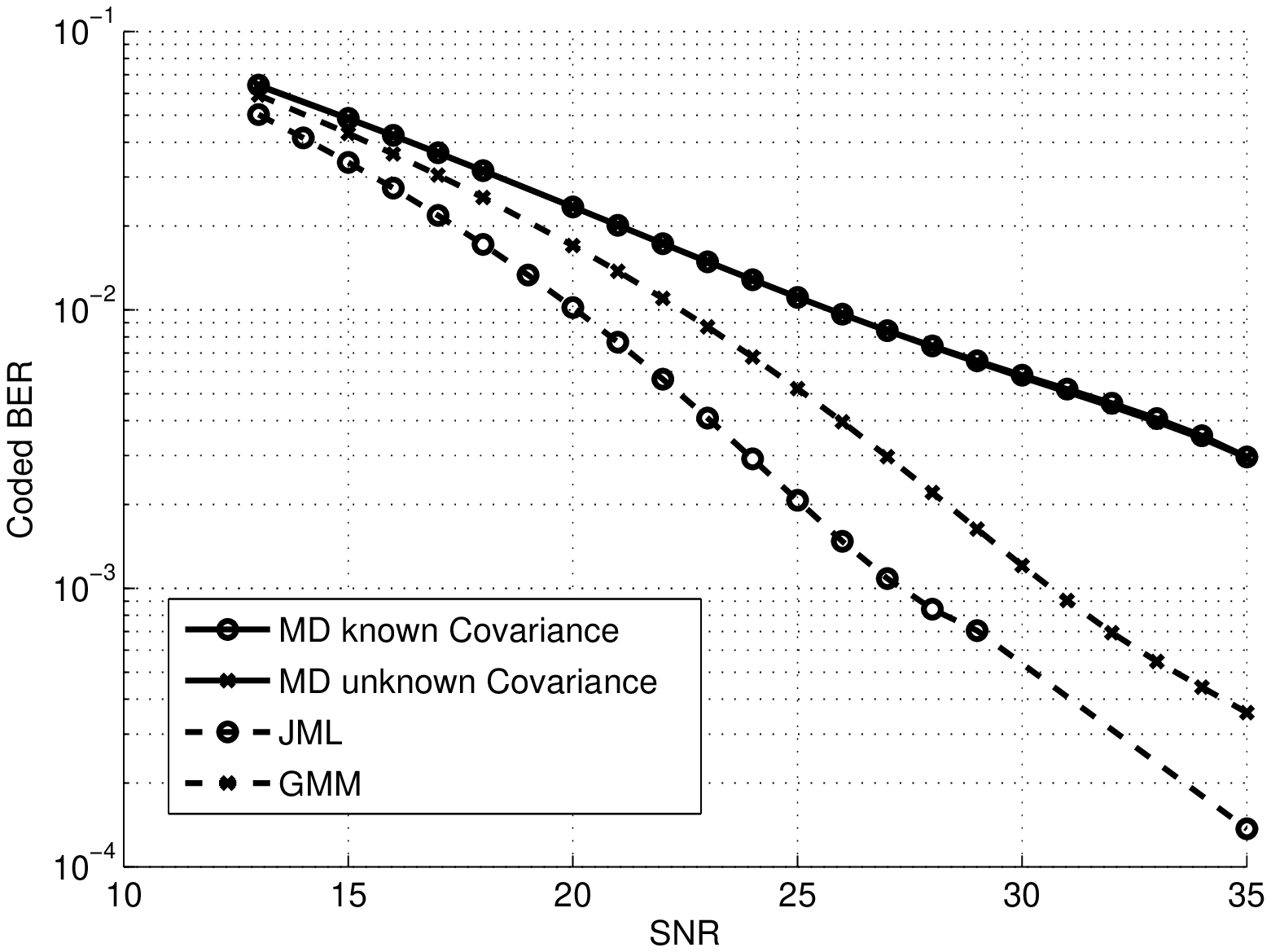}
\caption{Coded BER: $K=3,\;L=4,\;d=2$}
\label{fig:C_K_3_M_4_d_2}
\end{minipage}
\ifCLASSOPTIONonecolumn
\fi
\end{figure*}
The BER performance of the MD Detector and the ML Detector are compared in this section. As in B-IA literature, we also do not assume any CSIT, and we consider the two cases: (i)~CSIR is available and (ii)~I-CSIR is not available. The BER performance is averaged over $500$ independent channel realizations, and the unitary matrices $\vec{U}_i$s are chosen arbitrarily, but held constant across a given channel realization. A single channel realization comprises of $500$ transmission symbols.
The noise variance per received dimension is $\sigma^2$, and $\text{SNR} = \sfrac{1}{\sigma^2}$. The channel values, $h_{i,j}$, are \textit{i.i.d}, and takes value from the complex Gaussian distribution with zero mean. The covariance of the channel is given as $\text{E}[h_{i,j} h_{i,j}^*] = \alpha_{i,j}$, where $\alpha_{i,j}$ represent the power with which the signal is received at the $i^{th}$ receiver from the $j^{th}$ transmitter. We consider the strong interference regime, where $\alpha_{i,i} = 1$, and $\alpha_{i,j} = 1, \text{ even }\forall i \neq j$. 

In Fig. \ref{fig:Un_K_2_M_3_d_2} and \ref{fig:C_K_2_M_3_d_2}, the uncoded and coded BER performances of $2-$user IC is provided, and the performances of $3-$user IC for the different detectors is compared in Fig. \ref{fig:Un_K_3_M_4_d_2} and \ref{fig:C_K_3_M_4_d_2}. Since the channel is held constant across $500$ received symbols, the estimated covariance for the MD detector is nearly the same as the actual covariance. Hence, the MD detector performance with a given covariance for the interfering signals and the estimated covariance are same for both $2-$user IC and $3-$user IC. However, the BER performance of ML detector with estimated interfering signal statistics is poorer than the ML detector with known CSIR. 
This performance degradation is due to the error in the pdf estimate of the interfering signal obtained using the EM algorithm.
The pdf estimate given by EM algorithm depends on the initialization\footnote{The initialization of the EM algorithm is based on a clustering algorithm. The algorithm finds the number of clusters present in the received samples, and the number of clusters is taken as the number of means in the GMM. The average of the sample values present in the cluster is taken as the initial mean value of that cluster.}, and that is the reason for this degradation. However, this degraded performance is still significantly better than the MD detector (even when the covariance of interfering signal in the MD detector is perfectly known). Hence, the preferred choice of detector in a practical scenario should be the ML detector. The trend in the coded BER\footnote{The code rate used in Figs. \ref{fig:C_K_2_M_3_d_2} and \ref{fig:C_K_3_M_4_d_2} is $\sfrac{1}{2}$, and the block length of the turbo code used in $512$.} performance curves is also very similar to uncoded BER performance (Fig. \ref{fig:Un_K_2_M_3_d_2} and \ref{fig:Un_K_3_M_4_d_2}), but the BER values are better because of the coding gain provided by turbo code. 

\begin{figure}[h]
\centering
\includegraphics[scale = 0.5]{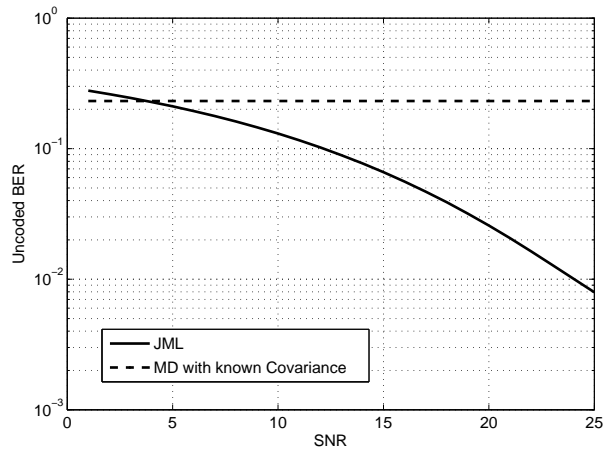}
\caption{Uncoded BER: $K=2,\;L=1,\;d=1$}
\label{fig:Un_K_2_M_1_d_1}
\ifCLASSOPTIONonecolumn
\fi
\end{figure}

In Figures \ref{fig:Un_K_2_M_3_d_2} thro \ref{fig:C_K_3_M_4_d_2}, there was no error floor in the BER performance, thereby satisfying the design constraint in (\ref{eqn:FIAConst}). In Fig. \ref{fig:Un_K_2_M_1_d_1}, we compare the BER performance for a $2-$user IC with SpAC value of one~($>~\sfrac{(L-K+1)}{L}$). The design constraint (\ref{eqn:FIAConst}) is based on the usage of MD detector. Hence, the BER performance floors when a SpAC value higher than $\sfrac{(L-K+1)}{L}$ is used. However, when the ML detector is used, there is no error floor, because the ML detector treats the $K$ transmitters as a single transmitters with more number of antennas, or equivalently, $|\mathcal{I}_i^{Rx}| = 0$. Hence any value of SpAC in the full range of $[0,1]$ is achievable.

\section*{Discussion}
\subsection{Application of B-IA vs B-FIA:}
\textit{B-IA:} The usage of B-IA in a Multi-User MIMO (MU-MIMO) scenario was given in \cite{Wang2011, Papadopoulos2011, Zhou2012}.  In \cite{Wang2011}, the transmit power in all the resources are made equal while having the structure of precoder for the B-IA scheme. It was shown that the B-IA scheme outperforms the Linear Zero-Forcing Beamforming scheme, and that too while enjoying a lesser overhead. In \cite{Papadopoulos2011}, the semi Blind IA was achieved by exploiting the nature of the channel power delay profile. In \cite{Zhou2012}, homogeneous block fading\footnote{All the channel values have same temporal correlation.} model was considered in a $2-$user BC, and a relative offset between coherence window of the two users was shown to achieve $\sfrac{4}{3}$ DoF. In general, to form the channel states as given in \cite{Jafar2012, Gou2011}, partial CSIT is needed; i.e., (i)~coherence interval \cite{Jafar2012, Zhou2012}, (or), (ii)~power delay profile of the channel \cite{Papadopoulos2011}, (or) (iii)
~reconfigurable antenna \cite{Gou2011, Wang2011}. For (i) and (ii), the transmitter needs the channel state information, and for (iii), the assumption was that the receiver has multiple antennas. However if all the receive antennas are used, higher DoF can be achieved even with spatially orthogonal transmission i.e., resource partitioning.

\textit{B-FIA:} Unlike the B-IA scheme, in the B-FIA there is no restriction on the channel states, and hence it is a practically feasible scheme. However, in order to use the detectors without I-CSIR the channel has to remain constant across a single block. In this work, the maximum achievable SpAC satisfying (\ref{eqn:FIAConst}) with no CSIT is obtained, and a precoder structure is provided to achievable the maximum SpAC. However, the constraint (\ref{eqn:FIAConst}) does not guarantee either a better BER, or a better rate. 
Hence, in order for B-FIA to be used in a practical scenario, the precoders are to be optimized to attain better BER or rate. In \cite{Arun2012}, a precoder structure similar to \eqref{eqn:ICBM_Matrices} was used, but with SpAC fixed to some arbitrary value, and the unitary matrices are replaced with a general $d \times d$ matrices which are designed by maximizing the minimum distance of the constellation seen at the receiver. In order to perform optimization, once again perfect CSIT is needed, but a robust precoder\footnote{A precoder set independent of SNR and SIR, but achieves better performance over a wide range of SNR and SIR} was chosen to benefit the low rate users. In \cite{Arun2012}, the LTE downlink was considered, and the MD detector with known covariance was used. Since the precoder design in \cite{Arun2012} is the optimized version of B-FIA, in this work we term the `Interference Cancelling Block Modulation' scheme in \cite{Arun2012} as Optimized B-FIA (OB-FIA). The numerical results in \cite{
Arun2012} shows that the LTE network with OB-FIA achieves better throughput than the network with no OB-FIA.
In \citep{Arun2012}, the MD detector with known covariance was used. Our current results in Figs. \ref{fig:Un_K_2_M_3_d_2} thro \ref{fig:C_K_3_M_4_d_2} indicate that the ML detector with estimated parameters\footnote{Note that the precoder structure enables the estimation.} performs much better than the MD detector. Hence, if ML detector (with estimated statistics) is used, much higher gain than reported in \cite{Arun2012} should be achievable.

\ifCLASSOPTIONonecolumn
\fi
\subsection{What is the usefulness of Table~\ref{tab:Different_SpACBCIC}, if ML detector can achieve $1$ SpAC?}
As argued in the section \ref{subsec:SpAC_CSIR_available} the ML detector with known CSIR can achieve $1$ SpAC. However, the usage of ML detector is limited because of: (i)~channel states of the interfering signal, and the modulation used by each transmitter should be known in-order to form the joint pdf of the interfering signal, and (ii)~the decoding complexity. The decoding complexity grows exponentially with the number of interfering signal, and even if any decoding complexity is allowed, obtaining I-CSIR needs a pilot design like in \cite{KVGHRG2014}, which enables the estimation of channel states of the interfering signal and the modulation used. When ML detector cannot be employed, the MD detector can be used. In such a case, the values provided in Table~\ref{tab:Different_SpACBCIC} are the maximum achievable SpAC.


\ifCLASSOPTIONonecolumn
\fi
\section{Conclusion} \label{sec:conclusion}
A novel Blind Fractional Interference Alignment (B-FIA) scheme was introduced in this work for both broadcast and interference channels. It was analytically shown that with no CSIT, the maximum achievable SpAC for the SISO was $\sfrac{(L-K+1)}{L}$, and for MIMO the maximum achievable SpAC was given by $\sfrac{(ML-K+1)}{ML}$ for BC, and $\sfrac{(ML-\frac{KM}{N}+1)}{ML}$ for IC. A simple precoder design was provided to achieve any value of SpAC lower than or equal to these values. The precoder design provided does not require I-CSIR for the SISO link. For I-CSIR to be known, all the transmitters should use pilot or control signals, which however will result in additional overhead. These overheads can be avoided in the CB-FIA scheme, where the interference statistics for the MD and ML detectors were estimated from the received samples by exploiting the precoder structure. Even though the precoders are designed based on the usage of the MD detector, numerical results indicate that the ML detector with estimated 
parameters should be preferably used at the receiver.

\ifCLASSOPTIONonecolumn
\fi
\appendix
\numberwithin{equation}{subsection}
\setcounter{equation}{0}
\subsection{Proof for Theorem \ref{Thm:samePDF}} \label{App:ThmsamePDF_Pf}
From \cite{Yang2013}, the joint probability distribution of $\vec{z}$ is given by,
\begin{equation}
\begin{array}{llll}
\text{f}(\vec{z}) = \frac{1}{|\mathcal{X}|^{2l}\pi \sigma^2} \sum\limits_{\vec{x}\in \mathcal{X}^{2l}} exp(-\frac{\parallel\vec{z}-\vec{U}\vec{x}\parallel^2}{\sigma^2})
\end{array},
\end{equation}
and the marginal distribution is given by,
\begin{equation}\label{eqn:MargPDF}
\begin{array}{llll}
\text{f}(z_i) &=& \frac{1}{|\mathcal{X}|^{2l}\pi \sigma^2} \sum\limits_{\vec{x}\in \mathcal{X}^{2l}} exp(-\frac{|z_i-\vec{u}_i^T \vec{x}|^2}{\sigma^2})\\
&=& \frac{1}{|\mathcal{X}|^{2l}} \sum\limits_{\vec{x}\in \mathcal{X}^{2l}} \mathcal{CN}(\vec{u}_i^T\, \vec{x},\sigma^2)
\end{array},
\end{equation}
where $\vec{U}^T = [u_1\; \cdots\; u_{2l}]$. Hence, in order to prove Theorem \ref{Thm:samePDF}, it is sufficient to prove that the set, 
\begin{equation}
\begin{array}{lll}
\mathcal{M}_i = \{\vec{u}_i^T\, \vec{x},\; \forall \vec{x}\in \mathcal{X} \},\;\; i = 1,\cdots 2l
\end{array},
\end{equation}
has the same elements irrespective of the $i^{th}$ row vector of the real unitary matrix, $\vec{U}$.

In Theorem \ref{Thm:samePDF}, $\vec{U}$ is of dimension $2l\times 2l$. The general structure for a real unitary matrix with $l=1$ is given by,
\begin{equation}
\begin{array}{lll}
\vec{U} = \begin{bmatrix}
\cos\theta & \sin\theta \\ -\sin\theta & \cos\theta
\end{bmatrix}
\end{array}.
\end{equation}
Let $\vec{x}=[x_i^{[1]}\;x_j^{[2]}]^T,\;\;i,j=1,\cdots, |\mathcal{X}|$, then,
\begin{equation}
\begin{array}{lll}
\mathcal{M}_1 = \{\cos\theta\,x_i^{[1]} + \sin\theta\,x_j^{[2]} ,\; \forall i,j=1,\cdots, |\mathcal{X}|\}
\end{array},
\end{equation}
and,
\begin{equation}
\begin{array}{lll}
\mathcal{M}_2 = \{\cos\theta\,x_j^{[2]} - \sin\theta\,x_i^{[1]} ,\; \forall i,j=1,\cdots |\mathcal{X}|\}
\end{array}.
\end{equation}
Since both $x^{i}$ and $x^{j}$ belongs to the same set $\mathcal{X}$, and using the property of the set $\mathcal{X}$ ($\mathcal{X} = \{\pm a_i,\; i=1,\cdots, \sfrac{|\mathcal{X}|}{2} \}$), the proof for Theorem~\ref{Thm:samePDF} is trivial for $l=1$, and the sets $\mathcal{M}_1$ and $\mathcal{M}_2$ also follow the property of the set $\mathcal{X}$, i.e., $\mathcal{M}_j= \{\pm m_i^{[j]},\; i=1,\cdots, \sfrac{|\mathcal{M}_j|}{2} \},\;j=1,2$.

From \cite{Bellman1970} and \cite{Lam2010}, the general structure of the real unitary matrix is given by,
\begin{equation}
\begin{array}{llll}
\vec{U}=\prod\limits_{\substack{i=1,\cdots, 2l \\ j=i+1,\cdots, 2l}} \vec{U}_{i,j}
\end{array},
\end{equation}
and,
\begin{equation}
\begin{array}{lll}
\vec{U}_{i,j}^{[\setminus i,\setminus j]} = \vec{I}_{2l-2},\quad \vec{U}_{i,j}^{[i,j]} = \begin{bmatrix}
\cos\theta_{ij} & \sin\theta_{ij} \\ -\sin\theta_{ij} & \cos\theta_{ij}
\end{bmatrix}
\end{array}
\end{equation}
where $\vec{U}_{i,j}^{[\setminus i, \setminus j]}$ represents the matrix formed by removing the $i^{th}$ and $j^{th}$ rows and also $i^{th}$ and $j^{th}$ columns from the matrix $\vec{U}_{i,j}$, and $\vec{U}_{i,j}^{[i,j]}$ is a $2\times 2$ matrix formed from the elements of $\vec{U}_{i,j}$ which are the intersection of $i^{th}$ and $j^{th}$ rows with $i^{th}$ and $j^{th}$ columns, and all other elements are zero. In short, any possible real unitary matrix is a product of $\sfrac{(2l)(2l-1)}{2}$ matrices (and hence, a function of $\sfrac{(2l)(2l-1)}{2}$ variables), and in each case a rotation is performed between two elements while keeping the rest of the elements remains unchanged. An example with $l=2$ is given below:
\ifCLASSOPTIONtwocolumn
\begin{equation}
\begin{array}{lll}
\vec{U} &= &\begin{bmatrix}
\cos\theta_{12} & \sin\theta_{12} & 0 & 0 \\ 
-\sin\theta_{12} & \cos\theta_{12} & 0 & 0\\ 
0 & 0 & \cos\theta_{34} & \sin\theta_{34}\\ 
0 & 0 & -\sin\theta_{34} & \cos\theta_{34}
\end{bmatrix} 
\\&& \times
\begin{bmatrix}
\cos\theta_{13} & 0 & \sin\theta_{13} & 0 \\ 
0 & \cos\theta_{24} & 0 & \sin\theta_{24}\\ 
-\sin\theta_{13} & 0 & \cos\theta_{13} & 0\\ 
0 & -\sin\theta_{24} & 0 & \cos\theta_{24}
\end{bmatrix}
\\&& \times
\begin{bmatrix}
\cos\theta_{14} & 0 & 0  & \sin\theta_{14}\\ 
0 & \cos\theta_{23} & \sin\theta_{23} & 0 \\ 
0 & -\sin\theta_{23} & \cos\theta_{23} & 0 \\
-\sin\theta_{14} & 0 & 0  & \cos\theta_{14}
\end{bmatrix}
\end{array},
\end{equation}
\else
\begin{equation}
\begin{array}{lll}
\vec{U} = \begin{bmatrix}
\cos\theta_{12} & \sin\theta_{12} & 0 & 0 \\ 
-\sin\theta_{12} & \cos\theta_{12} & 0 & 0\\ 
0 & 0 & \cos\theta_{34} & \sin\theta_{34}\\ 
0 & 0 & -\sin\theta_{34} & \cos\theta_{34}
\end{bmatrix} 
&& \times
\begin{bmatrix}
\cos\theta_{13} & 0 & \sin\theta_{13} & 0 \\ 
0 & \cos\theta_{24} & 0 & \sin\theta_{24}\\ 
-\sin\theta_{13} & 0 & \cos\theta_{13} & 0\\ 
0 & -\sin\theta_{24} & 0 & \cos\theta_{24}
\end{bmatrix}
\\&& \hspace{-40mm} \times
\begin{bmatrix}
\cos\theta_{14} & 0 & 0  & \sin\theta_{14}\\ 
0 & \cos\theta_{23} & \sin\theta_{23} & 0 \\ 
0 & -\sin\theta_{23} & \cos\theta_{23} & 0 \\
-\sin\theta_{14} & 0 & 0  & \cos\theta_{14}
\end{bmatrix}
\end{array},
\end{equation}
\fi
where we have combined $\vec{U}_{1,4}$ and $\vec{U}_{2,3}$ as a single matrix, and a similar approach can be for the other two matrices.
In any optimization problem, since all the elements of $\vec{x}$ takes value from the set $\mathcal{X}$, for $l=2$ case, $\theta_{14}$ will be given the same value as $\theta_{23}$ (since there is no priority among the elements of $\vec{x}$), and, from the $l=1$ case, the resultant set will also have the same property as that of the set $\mathcal{X}$. Once again, all the elements, after getting multiplied by $\vec{U}_{1,4}$ and $\vec{U}_{2,3}$, will take values from the same set. Proceeding similarly, it can be verified that in the optimization problem $\theta_{1,2}$ and $\theta_{1,3}$ will be set equal to $\theta_{3,4}$ and $\theta_{2,4}$ respectively. Yet again using the result of $l=1$ it can be shown that the resultant set,
\begin{equation}
\begin{array}{lll}
\mathcal{M}=\mathcal{M}_i = \{\vec{u}_i^T\, \vec{x},\; \forall \vec{x}\in \mathcal{X} \},\;\; i = 1,\,2,\,3,\,4
\end{array},
\end{equation}
and the optimization problem will be a function of three variables ($\theta_{1,2},\,\theta_{1,3},\,\theta_{1,4}$). This procedure can be extended for any value of $l$, and in each case the optimization problem will be a function of $(2l-1)$ variables, $\theta_{1,j},\;j=2,\cdots,2l-1$, and 
\begin{equation}
\begin{array}{lll}
\mathcal{M}=\mathcal{M}_i = \{\vec{u}_i^T\, \vec{x},\; \forall \vec{x}\in \mathcal{X} \},\;\; i = 1,\cdots 2l
\end{array},
\end{equation}
which completes the proof for Theorem \ref{Thm:samePDF}.
\begin{definition}[Useful real unitary matrix]
In general, the matrix $\vec{U}$ will be generated from some optimization problem, and based on the set $\mathcal{X}$, it can be shown that the optimization problem will be function of $(2l-1)$ variables instead of $\sfrac{(2l)(2l-1)}{2}$ (for any real unitary matrix). Hence, we refer to the collection of these matrices, which are a function of $2l-1$ variables, as useful real unitary matrices.
\end{definition}




\subsection{Covariance estimation}\label{App:covEst}
In practical scenarios, the covariance estimation is performed on the pilot locations, where the desired pilot data are known. The covariance of interference plus noise term is estimated by performing average across the resource elements (for block fading, or slow fading channels), and the estimated covariance is given by,
\begin{equation}\label{app:eqn:cov_est}
\begin{array}{lll}
\vec{R} = \sum\limits \parallel\vec{y}-\vec{H}\vec{x}_{pilot}\parallel^2
\end{array}.
\end{equation}
This averaging is performed only over the pilot locations, because in the data locations the desired signal, $\vec{x}$, is unknown. The structure in (\ref{eqn:ICBM_Matrices}) has a resource element in which the desired signal is not present, or can be treated as desired signal is known, and it takes the value of zero. Hence, as in \textit{pdf} estimation, these resource elements can be used to estimate the covariance $\vec{R}$ (using~(\ref{app:eqn:cov_est})).



\end{document}